\documentclass[journal,comsoc]{IEEEtran}%
\usepackage[T1]{fontenc}
\usepackage{multirow}
\usepackage[latin9]{inputenc}
\usepackage[english]{babel}
\usepackage[table]{xcolor}
\usepackage{caption}
\usepackage{collcell}
\usepackage{hhline}
\usepackage{url}
\usepackage{pgf}
\usepackage{graphicx}
\def\colorModel{hsb} 
\newcommand\ColCell[1]{
  \pgfmathparse{#1<50?1:0}  
    \ifnum\pgfmathresult=0\relax\color{white}\fi
  \pgfmathsetmacro\compA{1}      
  \pgfmathsetmacro\compB{#1/100} 
  \pgfmathsetmacro\compC{1}      
  \edef\x{\noexpand\centering\noexpand\cellcolor[\colorModel]{\compA,\compB,\compC}}\x #1
  } 
\newcolumntype{E}{>{\collectcell\ColCell}m{0.4cm}<{\endcollectcell}}  

\usepackage{color, colortbl}
\definecolor{lightGray}{gray}{0.9}
\definecolor{darkGray}{gray}{0.5}
\definecolor{white}{gray}{0.99}
\usepackage{tikz}
\usetikzlibrary{shapes.geometric, arrows}
\ifCLASSINFOpdf
\else
\fi
\usepackage{amsmath}
\usepackage[cmintegrals]{newtxmath}
\ifCLASSOPTIONcompsoc
  \usepackage[caption=false,font=normalsize,labelfont=sf,textfont=sf]{subfig}
\else
  \usepackage[caption=false,font=footnotesize]{subfig}
\fi
\usepackage{booktabs}
\usepackage{colortbl}
\usepackage{subfig}
\begin{document}
\title{Using Channel State Information for Physical Tamper Attack Detection in OFDM Systems: A~Deep Learning Approach}%
\markboth{}{Shell \MakeLowercase{\textit{et al.}}: Using Channel State Information for Physical Tamper Attack Detection in OFDM Systems: A Deep Learning Approach Journals}%
\author{Eshagh~Dehmollaian,~Bernhard~Etzlinger,~N\'uria~Ballber~Torres,~Andreas~Springer
Institute~for~Communications~Engineering~and~RF-Systems\\
Johannes~Kepler~University,
Linz,~Austria
\thanks{This work has been supported in part by the SCOTT project, Johannes Kepler University, Linz Institute of Technology (LIT) under grant no. LIT-2016-1-SEE-026, and the InSecTT project. SCOTT and InSecTT have received funding from the Electronic Component Systems for European Leadership Joint Undertaking under grant agreement No 737422 and No 876038, respectively. This Joint Undertaking receives support from the European Union's Horizon 2020 research and innovation programme and Austria, Spain, Finland, Ireland, Sweden, Germany, Poland, Portugal, Netherlands, Belgium, Norway.}%
}%
\maketitle%
\begin{abstract}
This letter proposes a deep learning approach to detect a change in the antenna orientation of transmitter or receiver as a physical tamper attack in OFDM systems using channel state information. We treat the physical tamper attack problem as a semi-supervised anomaly detection problem and utilize a deep convolutional autoencoder (DCAE) to tackle it. The past observations of the estimated channel state information (CSI) are used to train the DCAE. Then, a post-processing is deployed on the trained DCAE output to perform the physical tamper detection. Our experimental results show that the proposed approach, deployed in an office and a hall environment, is able to detect on average 99.6\% of tamper events (TPR = 99.6\%) while creating zero false alarms (FPR = 0\%).
\end{abstract}
\IEEEpeerreviewmaketitle
\section{Introduction}
\IEEEPARstart{I}{}f wireless networks are used in critical infrastructures such as airports, military installations, etc., a high level of security is required. Among the many different threats in wireless networks, we consider physical tampering with a device and more specifically the movement or relocation of static transmitters or receivers. Such an attack impacts the functionality of a wireless network by changing the radio channel. An example, which is investigated by \cite{Bagci}, is a surveillance system using WiFi-based cameras integrated with antennas to monitor critical infrastructure. Another example is RF fingerprint-based localization systems using PHY-layer aspects (e.g., \cite{csi_deep}). In both cases, the locations of transmitters need to be fixed in order for the system to work properly. If the transmitters are moved, with high probability the functionality of the systems is destroyed. Therefore, a physical tamper attack detection mechanism is needed.

In order to solve this issue, we use PHY layer information already available in wireless networks. Previously, Fabia et al. \cite{Fabia} utilized RSSI information to detect identity-based attacks in wireless networks. Patwari et al. \cite{Patwari} exploited channel impulse response (CIR) information to define link signatures for transmitter location distinction. Bagci et al. \cite{Bagci} proposed an algorithm based on the channel state information (CSI) values of a transmission at multiple receivers for solving the physical tamper attack problem. These works consider direct seqeuence spread spectrum \cite{Patwari} or OFDM \cite{Bagci,Fabia} based systems.

Since OFDM is currently the most common transmission technology \cite{Ramjee}, we follow \cite{Bagci,Fabia} and use CSI (which contains more information than RSSI) in an OFDM-based wireless system. The direct relation of CSI to the radio channel makes it possible to detect physical changes due to a tamper attack. Besides, the radio channel is also sensitive to other changes in the environment, like movement of people and/or objects, which are clearly not related to an attack. Thus, an attack detection algorithm must be able to distinguish those~cases.

To differentiate between environmental changes and tamper attacks on the transmitter, \cite{Bagci} suggests to use multiple receivers and calculate different distances between the CSI of the offline and the online phase to detect the physical tamper attack. This approach is based on the assumption, that environmental variations (e.g., from moving persons) will not affect all receivers.
Unlike \cite{Bagci}, instead of calculating the distances, we propose to use machine learning methods to detect the physical tamper attack based on CSI at a receiver. The main contributions of this letter~are:
\begin{itemize}
  \item We exploit a machine learning approach for physical tamper attack detection. We consider the problem as a semi-supervised anomaly detection problem.
  \item We use a deep convolutional autoencoder (DCAE) with a post-processing unit to learn CSI-characteristics in tamper free scenarios and use it to detect physical tamper attacks.
  \item For a robust tamper detection algorithm, we propose to use a probability density~function (pdf) approximation of the DCAE reconstruction error. The detection performance is compared with existing work in the literature.
\end{itemize}

The letter is organized as follows: The tamper detection framework is introduced in section~II. The operational phases and the experimental results are presented in Sections III and IV, respectively. Finally, Section~V concludes~the~letter.
\section{Tamper Detection Framework}
\subsection{Problem Statement}
In wireless communication the transmitted signal is modified by the mobile radio channel. This channel is determined by the propagation of the electromagnetic waves from transmitter to receiver, which in turn is defined by the surrounding environment. Any change in the environment also changes the mobile radio channel. Thus, the main difficulty in detecting a physical tamper attack in form of movement or relocation of transmitters or receivers is to distinguish between changes caused by the physical tamper attack and changes caused by modifications in the surrounding like persons walking by. Our goal is to use a machine learning approach applied to the magnitude of the estimated CSI ($|\mathbf{\hat{H}}|$\footnote{The estimated signal taken from the receiver is denoted by $\mathbf{\hat{H}}$.})\cite{abdolla} that detects tamper attacks on the transmitter despite changes in the environment.

We formulate the problem of detecting the physical tamper attack as a data-driven semi-supervised anomaly detection problem. According to \cite{anomaly_survey}, semi-supervised anomaly detection refers to the problem of finding patterns in data that do not conform to expected behavior. For the tamper detection problem we aim to detect if the characteristics of the magnitude of CSI does not conform to the characteristics learned~in~tamper~free~scenarios.

In order to investigate the performance of the proposed method in the physical tamper attack problem, three different methods are taken into account as below:
\subsection{Conventional Threshold Detection}
Similar to \cite{Bagci}, a simple and straightforward approach is to use a distance metric followed by a threshold decision. If the distance is greater than a certain threshold, the algorithm will detect tampering. $\mathbf{H}_\mathrm{Off}$ is the collected magnitude of the estimated CSI while the transmitter is in a tamper free scenario (Offline phase) given by:
\begin{equation}
\mathbf{H}_\mathrm{Off}\triangleq[|\mathbf{\hat{H}}_{\mathrm{Off}}^1|,|\mathbf{\hat{H}}_{\mathrm{Off}}^2|,\dots, |\mathbf{\hat{H}}_{\mathrm{Off}}^{{N_{\mathrm{Off}}}}|]^T\in \mathbb{R}^{{N_{\mathrm{Off}}\times Sc }}\,.
\label{alg1}
\end{equation}
It is used for comparison with the newly received magnitude of the estimated CSI $\mathbf{H}_\mathrm{On}$ (Online phase) represented as
\begin{equation}
\mathbf{H}_\mathrm{On}\triangleq[|\mathbf{\hat{H}}_{\mathrm{On}}^1|,|\mathbf{\hat{H}}_{\mathrm{On}}^2|,\dots, |\mathbf{\hat{H}}_{\mathrm{On}}^{{N_{\mathrm{On}}}}|]^T\in \mathbb{R}^{{N_{\mathrm{On}}\times Sc}}\,.
\label{alg2}
\end{equation}
A distance metric and a threshold decision is calculated as,
\begin{equation}
\begin{aligned}
\mathbf{D}_{i,j}\triangleq Distance(|\mathbf{\hat{H}}_{\mathrm{Off}}^i|,|\mathbf{\hat{H}}_{\mathrm{On}}^j|)\qquad ,i &=1,\dots,N_{\mathrm{Off}} \\
,j &=1,\dots,N_{\mathrm{On}}
\label{alg3}
\end{aligned}
\end{equation}
\begin{equation}
{mean}_{\forall i,j}(\mathbf{D}_{i,j}) \mathop{\lessgtr}_{Tampering}^{Tamper~Free} Threshold \,,
\label{alg4}
\end{equation}
where $Sc$ is the number of subcarrier, $N_{\mathrm{Off}}$ and $N_{\mathrm{On}}$ are the number of frames in the offline phase and online phase, respectively. As in \cite{Bagci}, $Distance$ in (\ref{alg3}) refers to a distance metric (e.g., $Euclidean~distance$). The distance metric quantifies the distance between two CSI vectors. Afterwards, the mean value of the distance is considered to make the decision. In this work, this approach is referred to as \textbf{Method~1}. As shown in Sec.~IV, this method has a poor attack detection performance.

\subsection{Using Deep Convolutional Autoencoder}
To detect modifications in the CSI, the DCAE can be applied \cite{cae}. It is a well-suited method for feature extraction and anomaly scoring in static environments. In what follows we first introduce the DCAE approach, before proposing a post-processing unit to enhance robustness.

DCAE is a method for representation learning that usually is used in image processing applications. It maps the input into a compressed representation space (i.e., latent space) with a number of convolutional layers and  max-pooling layers through the encoding procedure from which the decoding part reconstructs the input data with a number of convolutional layers and upsampling layers. The training is performed by minimizing the difference between input data and reconstructed data. Since a DCAE utilizes convolutional layers, it adopts local information to reconstruct the signal. The advantage of using the DCAE is that it automatically detects the most important characteristics of the training data and outputs a scoring value on how well new input data fits to those characteristics.
\begin{figure}
\centering
{\def\svgwidth{0.3\textwidth}
\fontsize{7pt}{9pt}\selectfont
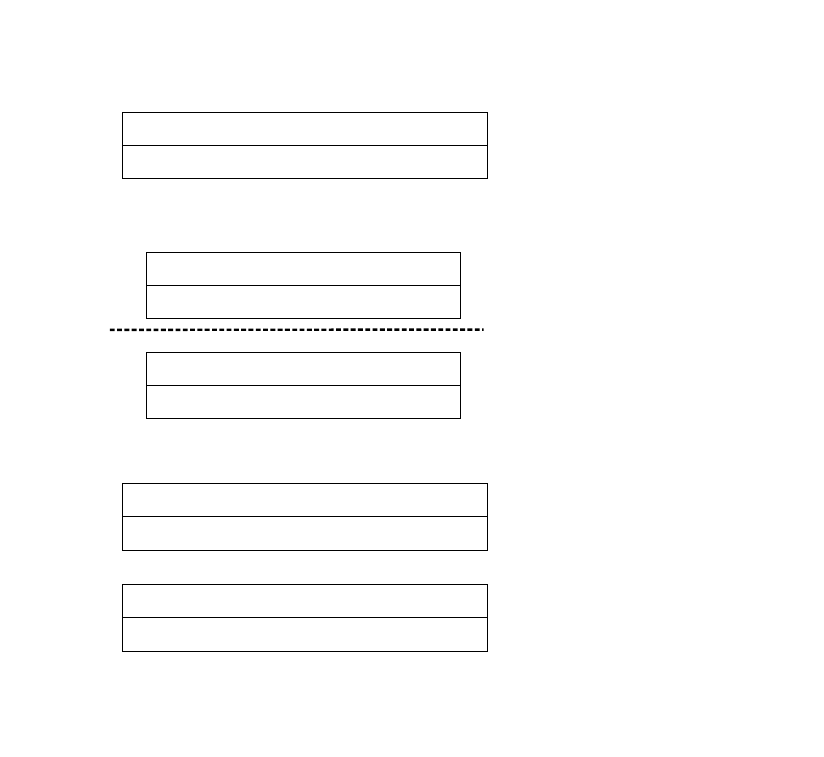}
\caption{DCAE to compare measured and reconstructed CSI, i.e. $|\mathbf{\hat{H}}|$ and $|\mathbf{\hat{H}|_{rec}}$, respectively. The reconstruction is based on the tamper free CSI characteristics that are encoded into the latent space.}
\label{DCAE}
\end{figure}

As shown in Fig.~\ref{DCAE}, the DCAE consists of two main stages: 1) The encoding procedure, which is a compression of the input data $|\mathbf{\hat{H}}|$ into a lower dimension space , the latent space. The compression is done with $n$ layers in the encoding procedure where each layer consists of a one-dimensional (1D) convolutional layer with $F$ filters, each with length $L$. The rectified linear unit (ReLU) activation function is used in the Conv1D layers. Then, a max pooling layer with parameter $M$ is appended. 2) The decoding procedure, which reconstructs the input data from the compressed representation ($|\mathbf{\hat{H}|_{rec}}$). For the decoding procedure, the structure of the encoding procedure is mirrored. At the end of the decoder, a fully connected layer, with the same number of neurons as the length of input data with sigmoid activation~function, is employed. The output of the DCAE is the reconstruction error vector $\mathbf{e}$ given by:
\begin{equation}
\mathbf{e}=|\mathbf{\hat{H}|_{rec}}-|\mathbf{\hat{H}}| \,.
\label{rec}
\end{equation}

The number of layers, number of filters, and size of filters were found by simulation experiments such that we achieved best performance while having moderate computational requirements, i.e. minimizing the reconstruction error of the training data while requiring low computation time.

Two methods that use the DCAE for the physical tamper attack problem are proposed as below:

\subsubsection{DCAE with distance threshold}
Instead of using $\mathbf{H}_\mathrm{Off}$ and $\mathbf{H}_\mathrm{On}$ in the previous approach, $\mathbf{E}_\mathrm{Off}$ and $\mathbf{E}_\mathrm{On}$ are used as,
\begin{equation}
\mathbf{E}_\mathrm{Off}\triangleq[\mathbf{e}_{\mathrm{Off}}^1,\mathbf{e}_{\mathrm{Off}}^2,\dots,\mathbf{e}_{\mathrm{Off}}^{{N}_{\mathrm{Off}}}]^T\in \mathbb{R}^{{N_{\mathrm{Off}}\times Sc}},
\label{eq8}
\end{equation}
\begin{equation}
\mathbf{E}_\mathrm{On}\triangleq[\mathbf{e}_{\mathrm{On}}^1,\mathbf{e}_{\mathrm{On}}^2,\dots,\mathbf{e}_{\mathrm{On}}^{{N}_{\mathrm{On}}}]^T\in \mathbb{R}^{{N_{\mathrm{On}}\times Sc}},
\end{equation}
where $\mathbf{e}_\mathrm{Off}^i$ and $\mathbf{e}_\mathrm{On}^j$ are the reconstruction error vectors
\begin{equation}
\mathbf{e}_\mathrm{Off}^i=|\mathbf{\hat{H}}_\mathrm{Off}^i|-|\mathbf{\hat{H}}_\mathrm{Off}^i|_{\mathrm{rec}}, \qquad i=1,\dots,N_\mathrm{Off},
\end{equation}
\begin{equation}
\mathbf{e}_\mathrm{On}^j=|\mathbf{\hat{H}}_\mathrm{On}^j|-|\mathbf{\hat{H}}_\mathrm{On}^j|_{\mathrm{rec}}, \qquad j=1,\dots,N_\mathrm{On}\,.
\end{equation}
In this work, this approach is referred to as \textbf{Method~2}.
\subsubsection{DCAE with pdf estimator}
To increase robustness in dynamic environments, we~consider an approximation of pdf of anomaly score in this approach which is referred to as \textbf{Method~3}. Similar to the method 2, $\mathbf{E}_\mathrm{Off}$ is considered as in (\ref{eq8}). The norm (i.e., the Euclidean norm) of the $\mathbf{E}_\mathrm{Off}$ is defined as~$Anomaly~Score$ $\mathbf{a}$ calculated~by:
\begin{equation}
\mathbf{a} =\parallel \mathbf{E}_\mathrm{Off} \parallel_2\epsilon \mathbb{R}^{N_{\mathrm{Off}}}\,.
\end{equation}
~~By computing the Euclidean norm of the output of the DCAE we obtain the anomaly scores of frames and use it to detect a physical tamper attack. As in many density-based anomaly detection approaches in the literature, to achieve a robust attack detector, the attack detection algorithm can either be a parametric or a non-parametric one. Since it is possible that the data does not fit well to any member of a parametric family of distributions, our approach is based on a pdf approximation of the anomaly score. We have used the Gaussian kernel density estimation from the Sklearn library \cite{sklearn} for the pdf estimator.

By measuring the similarity between the pdf approximations of the anomaly scores in the offline (where we have only tamper free data) and online phase, a physical tamper attack can be detected. In other words, if the newly estimated $|\mathbf{\hat{H}}|$ is sufficiently similar to the tamper free data from the offline phase, the DCAE is able to well reconstruct $|\mathbf{\hat{H}}|$ and the reconstruction error is small. Otherwise, the error is large.

We decide if a tamper attack has happened by measuring the distance between the stored pdf in the database and the obtained pdf in the online phase. If the distance exceeds a threshold, then the system triggers an attack detection alarm.

In order to measure the distance of two pdfs, we used the overlapping index \cite{overlappingindex}. The overlapping index $\eta:~\mathbb{R}^n~\times~\mathbb{R}^n\rightarrow~[0,1]$ is defined as:
\begin{equation}
\begin{split}
&\eta\left(f_{Y_{\mathrm{Off}}}(a) ,f_{Y_{\mathrm{On}}}(a)\right)=\int_0^\infty \min\left\{f_{Y_{\mathrm{Off}}}(a) ,f_{Y_{\mathrm{On}}}(a)\right\}\, \text{d}a,
\end{split}
\label{dist}
\end{equation}
where $ f_{Y_{\mathrm{Off}}}(a)$ and $f_{Y_{\mathrm{On}}}(a)$ are the pdf of the anomaly score in the offline and online phase, respectively.
\section{Operational Phases}
We train and deploy the DCAE, as a state-of-the-art semi-supervised learning method for anomaly detection, and a post-processing unit applied to the physical tamper attack detection problem. Our method includes an offline and an online phase.

\subsection{Offline Phase}
\begin{figure}[t]
\centering
{
   \def\svgwidth{0.34\textwidth}   
   \fontsize{7pt}{9pt}\selectfont
   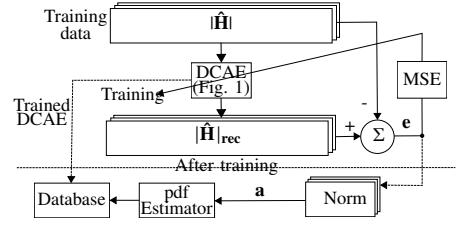
}
\caption{Offline phase: training the DCAE and storing the trained model and the pdf approximation of $\mathbf{a}$ in the database.}
\label{ML_off}
\end{figure}

Figure.~\ref{ML_off} shows the structure of the offline phase in which the tamper free CSI observations (denoted by $|\mathbf{\hat{H}}|$) from different environmental conditions including movement of people and static environments at different times are collected. 

These CSI measurements are utilized as the training data to train the DCAE. The DCAE weights are trained by using the mean square error (MSE) of the reconstruction error vector (\ref{rec}). Therefore, the MSE is used as loss function and the adam optimizer \cite{adam} was utilized to train the DCAE. We have made use of the Keras library \cite{keras} to build the proposed DCAE with 20~epochs~and~a~batch size of 100.

After training the DCAE, as described in method 3, the anomaly scores of $N_{\mathrm{Off}}$ frames are obtained. Finally, the weights of the trained DCAE and the pdf approximation of the anomaly score are stored in the database for the following use in the~online~phase.
\subsection{Online Phase}
\begin{figure}[t]
\centering
{
   \def\svgwidth{0.30\textwidth}   
   \fontsize{7pt}{9pt}\selectfont
   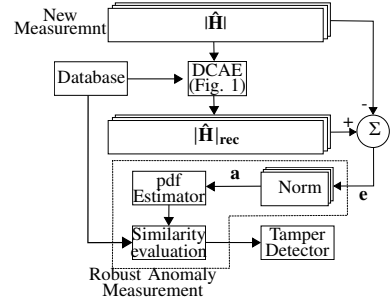
}
\caption{Online phase: using the trained DCAE and calculating and comparing the new $\mathbf{a}$ with the stored $\mathbf{a}$ in the database.}
\label{ML_on}
\end{figure}
As shown in Fig.~\ref{ML_on}, the new estimated CSI ("New measurement", $|\mathbf{\hat{H}}|$) is fed into the DCAE (with the weights that are loaded from the database). We find the anomaly score by computing the Euclidean norm of the output of the DCAE. We estimate the pdf of the anomaly score based on $N_{\mathrm{On}}$ successive frames. Finally, we evaluate the similarity of the obtained pdf with the one stored in the database. 
\section{Experimental Results}
\subsection{OFDM System}
To experimentally verify our proposed tamper attack detection method, we performed measurements using the Gnuradio OFDM project \cite{gnuradio}. The transmitter and receivers were realized via a USRP X310, equipped with a directional antenna \cite{antenna}, connected to a host computer for signal processing, using 200 data subcarriers, 8 pilot subcarriers, and 48 null subcarriers resulting in a channel bandwidth of 25~MHz. The OFDM symbol duration was 11.52 $\mu$s with 10.24 $\mu$s IFFT period and 1.28 $\mu$s guard interval. The carrier frequency was 2.55 GHz. A frame consisted of nine data OFDM symbols and three preamble symbols, Two preamble symbols were used for synchronization and one to estimate $\mathbf{\hat{H}}$ using a least squares approach. The amplitude of the estimated $\mathbf{\hat{H}}$ was~then~used~in~the~DCAE.

\subsection{Environment and Physical Tamper}
\begin{figure}[!ht]
   \includegraphics[width=0.23\textwidth]{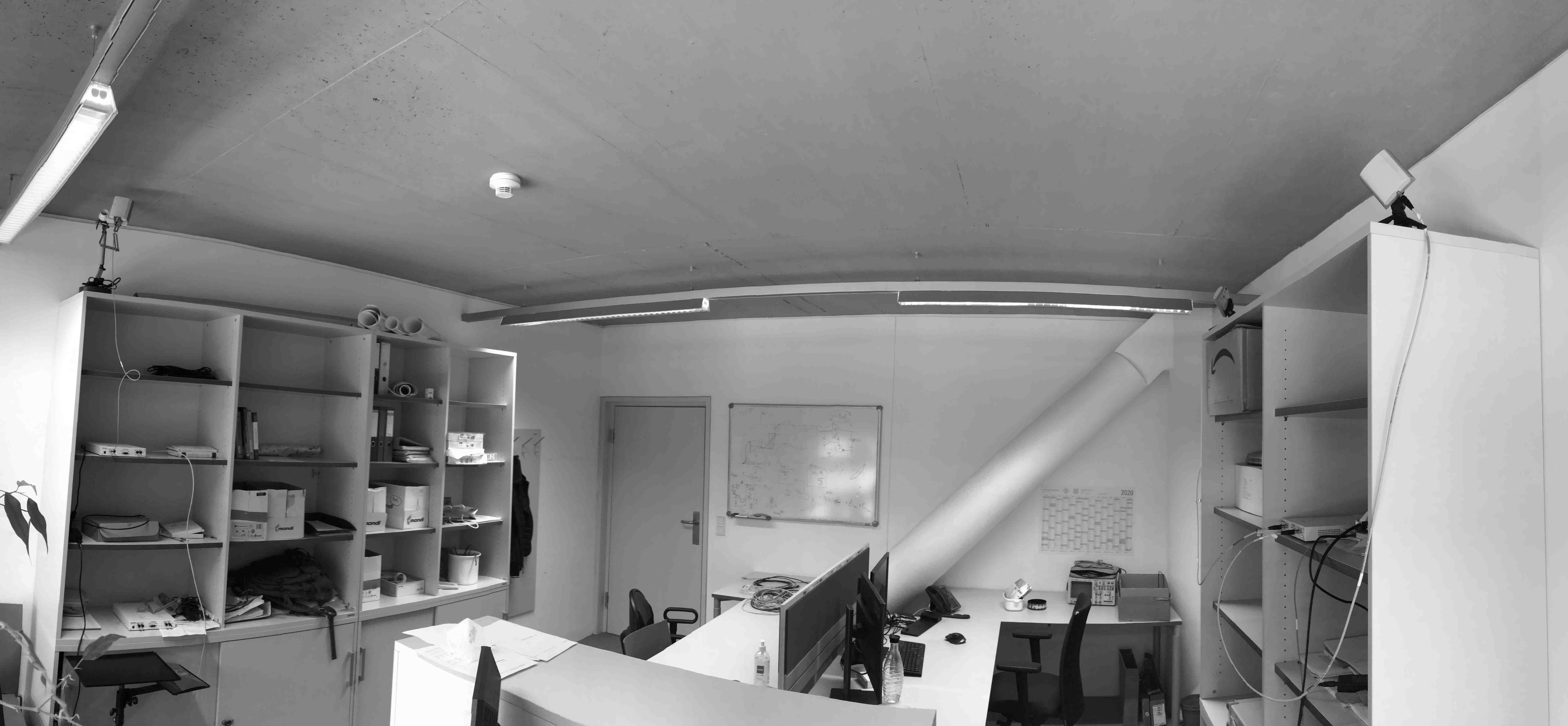} 
   \includegraphics[width=0.23\textwidth]{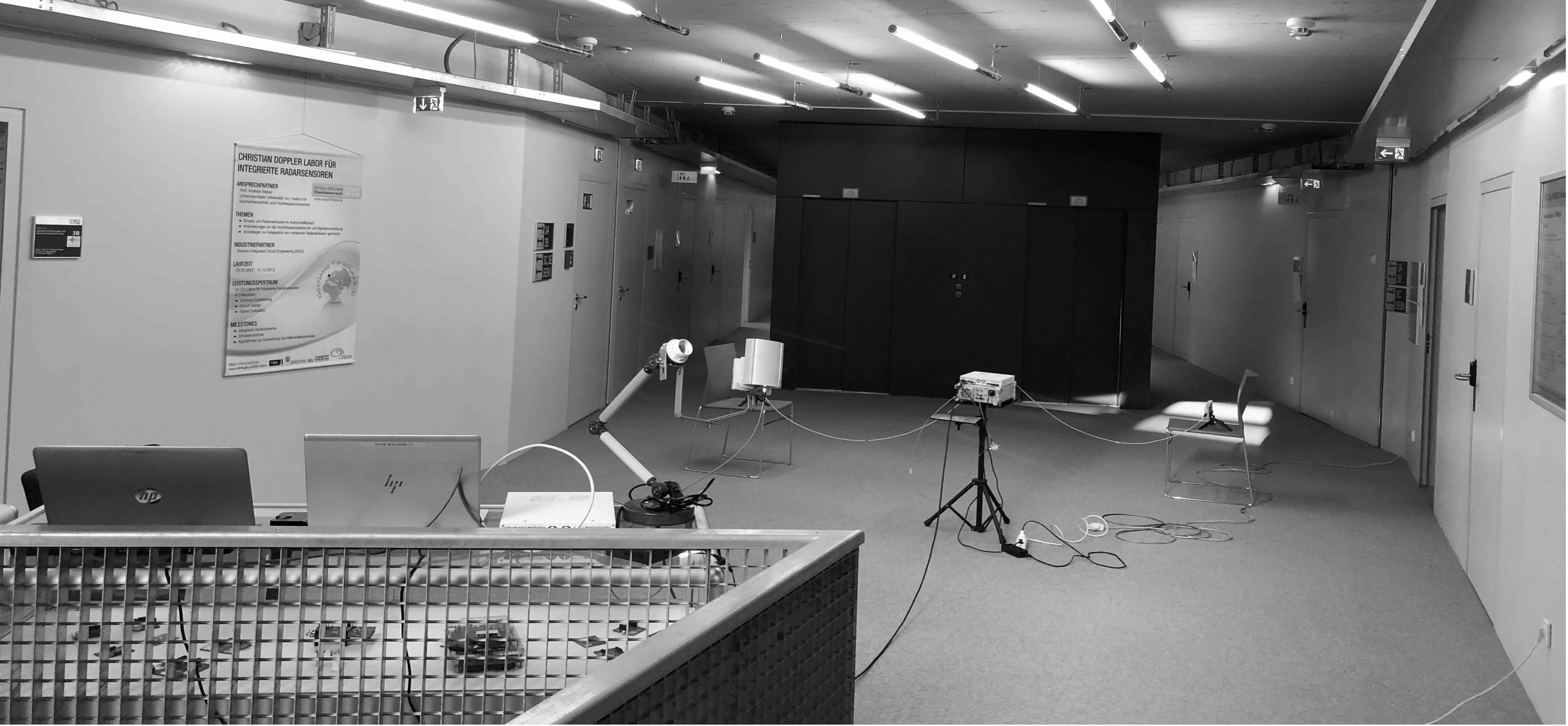}
     \vspace{1mm}
     \subfloat[\label{subfig-2:dummy}]{%
\def\svgwidth{0.19\textwidth}  
   \fontsize{7pt}{9pt}\selectfont
     \hspace{1mm}
     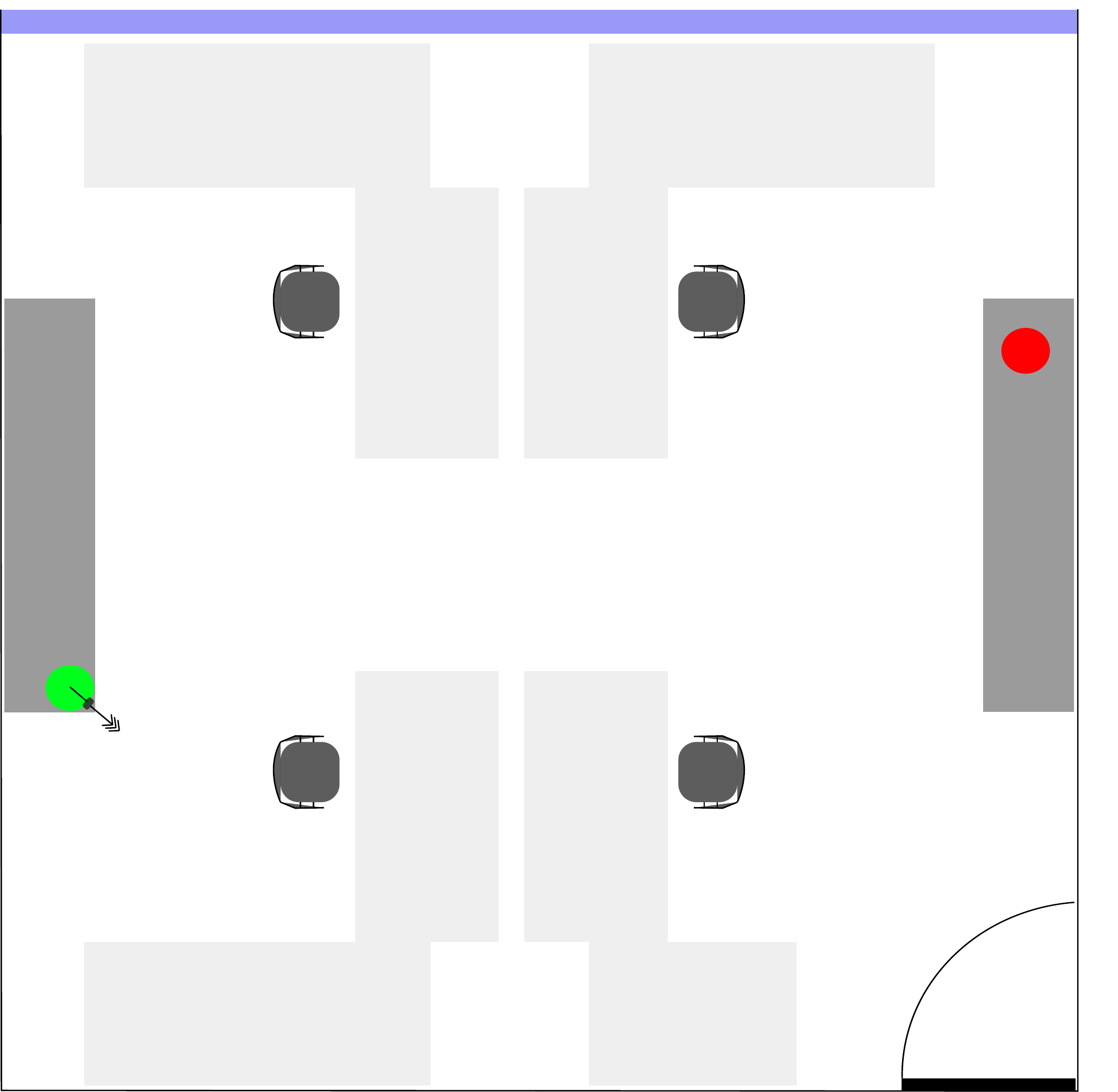
     }
    \hspace{3mm}
     \subfloat[\label{subfig-2:dummy}]{%
   \def\svgwidth{0.25\textwidth}   
   \fontsize{7pt}{9pt}\selectfont
     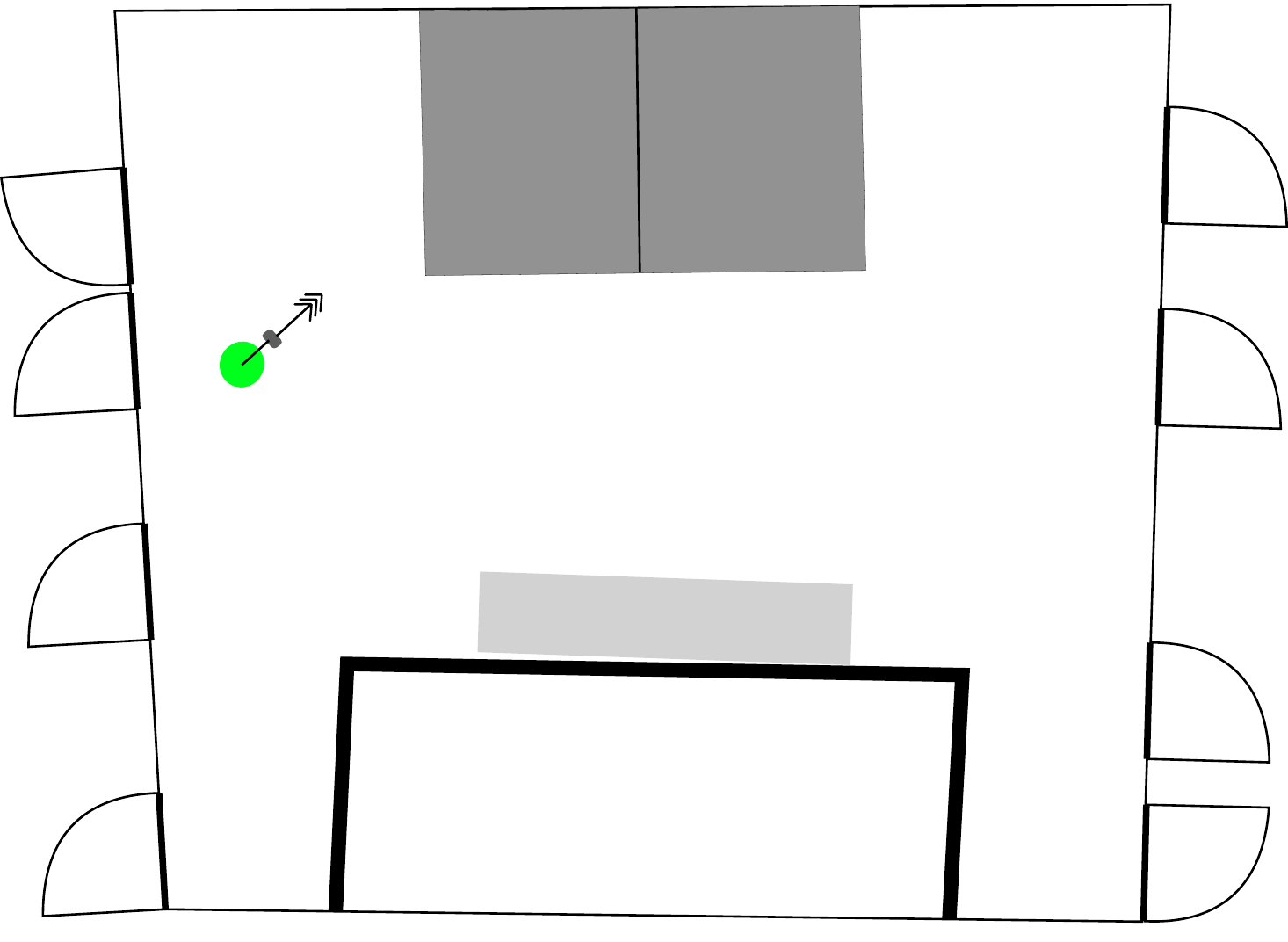
     }
     \caption{Measurement environments: (a) Office ($4\times6m^2$) and (b) Hall ($7\times12m^2$), depicted with photos and layout (not to scale). Orientations r1, r2, ... , r7~are considered as physical tamper attacks.}
     \label{fig:layouts}
   \end{figure}

We evaluated the aforementioned tamper detection methods in two different environments which are an office and a hall as depicted in Fig.~\ref{fig:layouts}. The transmitter (denoted by TX) and receivers (RX 1 and RX 2) were placed on top of the shelves with an elevation of 230~cm in the office environment and the desks with an~elevation~of~140~cm in the hall environment. RX~2 was considered only for the method from \cite{Bagci} which was used for comparison.

We considered 8 different antenna orientations, i.e., the tamper free default orientation and rotations r1, r2, ... r7 (c.f. Fig.~\ref{fig:layouts}) as physical tamper attacks.

\subsection{Experiment Methodology}
We considered 7 scenarios for the tamper free default orientation in the office environment, i.e., (A) a person sits on chair 1, (B) same as A one hour later, (C) a person walks in the area randomly, (D) same as C one hour later, (E) two persons walk in the area randomly, (F) same as E one hour later, (G) three persons walk in the area randomly. A, C, E, and G were considered in the hall environement.
\subsection{DCAE parameters}
Table \ref{table} summarized the parameters used to train the DCAE in Fig.~\ref{DCAE}. For the training data set a measurement set of ${N_{\mathrm{Off}}=40000}$ CSIs was collected. We trained the DCAE in which the default antenna orientation was used in all different scenarios. We trained the DCAE 5000 times with~different~initial weights. To compare the performance of DCAEs with different number of layers, two different DCAEs (DCAE1 [6~layers] and DCAE2 [8 layers]) were considered.

\begin{table}[t]
\centering
\setlength{\extrarowheight}{0pt}
\addtolength{\extrarowheight}{\aboverulesep}
\addtolength{\extrarowheight}{\belowrulesep}
\setlength{\aboverulesep}{0pt}
\setlength{\belowrulesep}{0pt}
\caption{DCAE parameters}
\tabcolsep=0.09cm
\begin{tabular}{c|c} 
\toprule
\rowcolor[rgb]{0.753,0.753,0.753} Description                                                                                                                                                                                               & Value                                                                                                                                                                                                      \\ 
\hline
Optimizer                                                                                                                                                                                                                                   & Adam                                                                                                                                                                                                       \\ 
\hline
Batch Size                                                                                                                                                                                                                                  & 100                                                                                                                                                                                                        \\ 
\hline
Number of Epochs                                                                                                                                                                                                                            & 20                                                                                                                                                                                                         \\ 
\hline
Learning Rate                                                                                                                                                                                                                               & 0.001                                                                                                                                                                                                      \\ 
\hline
DCAE1=$\begin{bmatrix}F_1 & L_1 & M_1 \\F_2 & L_2 & M_2 \\F_3 & L_3 & M_3 \end{bmatrix}$ & $\begin{bmatrix}10 & 52 & 2 \\10 & 26 & 2\\ 10 & 1 & 2\end{bmatrix}$                                             \\ 
\hline
DCAE2=$\begin{bmatrix}F_1 & L_1 & M_1 \\F_2 & L_2 & M_2 \\F_3 & L_3 & M_3\\ F_4 & L_4 & M_4\end{bmatrix}$ & $\begin{bmatrix}10 & 104 & 2 \\10 & 52 & 2 \\10 & 26 & 2\\ 10 & 1 & 2\end{bmatrix}$ \\
\bottomrule
\end{tabular}
\label{table}
\end{table}
\subsection{Tamper Detection Performance}
\begin{figure}[!htb]
    \begin{center}
        \scalebox{0.42}{\input{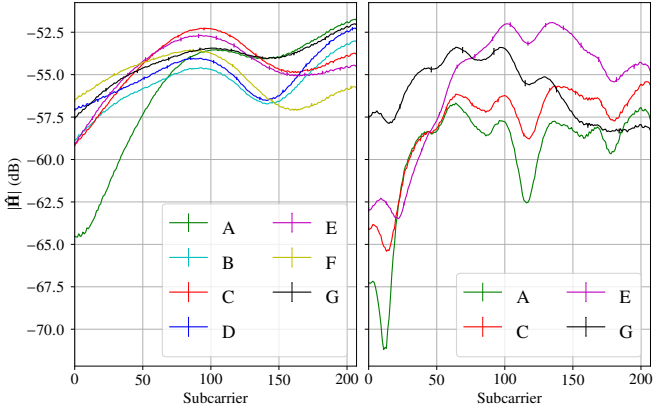}}
    \end{center}
    \caption{$|\mathbf{\hat{H}}|$ in tamper free scenarios presented in Sec.~IV.C in the office (left) and in the hall environement~(right).}
    \label{fig1}
\end{figure}

\begin{figure}[t]
    \begin{center}
        \scalebox{0.42}{
        \input{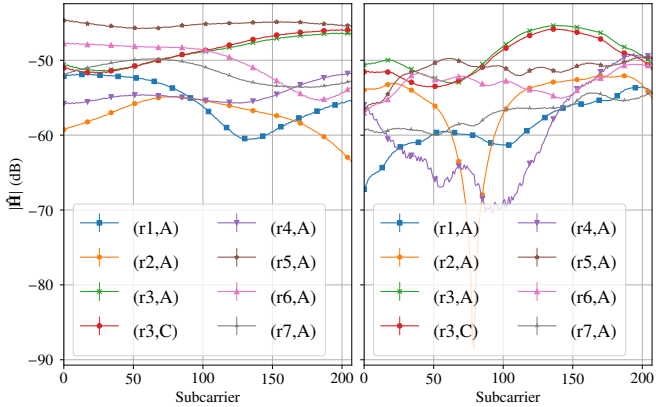}}
    \end{center}
    \caption{$|\mathbf{\hat{H}}|$ in tamper attack scenarios in the office (left) and in the hall environement (right).}
    \label{fig2}
\end{figure}

Figures \ref{fig1} and \ref{fig2} depict $|\mathbf{\hat{H}}|$ for tamper free and tamper attack scenarios in the both environments, respectively. In Fig.~\ref{fig1}, $|\mathbf{\hat{H}}|$ is shown for different scenarios with mean and variance depicted with the error bars over $N_{\mathrm{Off}}$ frames. From Fig.~\ref{fig1} it is obvious that the shape of $|\mathbf{\hat{H}}|$ is similar in all scenarios. Even if two people are walking, the shape of $|\mathbf{\hat{H}}|$ remains largely the same as compared to scenario (A). As can be seen, variations due to movement (two persons walking vs. one person sitting) only lead to slightly increased variance indicated by the error~bars in the insert of Fig.~\ref{fig1}.

\begin{figure}[t]
    \begin{center}    
        \scalebox{0.5}{\includegraphics{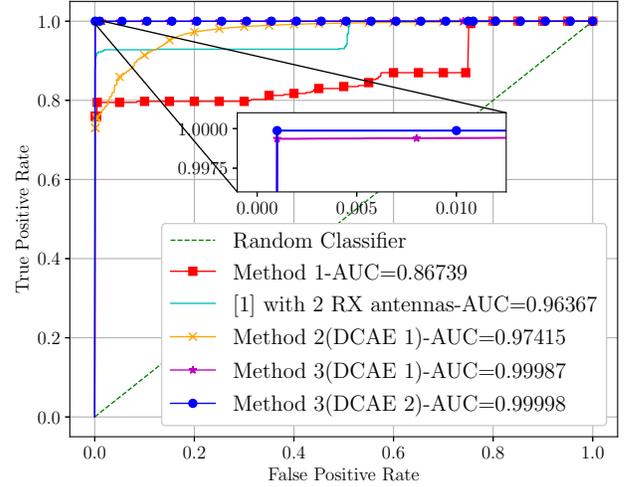}}
    \end{center}
    \caption{Comparison of the ROC curves as a function of the decision threshold. AUC is the area under the ROC Curve.}
    \label{roc}
\end{figure}

In contrast, in Fig.~\ref{fig2} the shape of $|\mathbf{\hat{H}}|$ changes when the orientation of the transmitter antenna is changed. As before, movement of people in the radio channel does not make a big difference in the shape of $|\mathbf{\hat{H}}|$ (c.f. (r3,A) and (r3,C) in Fig.~\ref{fig2}). We therefore conclude, that $|\mathbf{\hat{H}}|$ is a suitable input to the proposed~DCAE for tamper detection.
To have a performance comparison between the three aforementioned methods, a measurement set of 30000 CSI vectors of the tamper free scenarios and 30000 CSI vectors of the tamper scenarios were collected as the test set. We implemented method~1~by considering the Euclidean distance as its distance metric based on the CSI measured at one receiver. To assess the detection performance of the DCAE and the influence of the post-processing, the receiver operating characteristics (ROC) are depicted in Fig.~\ref{roc}. It is the average detection performance over the two environments. The ROC plots the true~positives~versus the false alarm rate by varying the threshold value.~While~the detection performance of method~1~is~poor,~its~advantage~is~simplicity. According to \cite{Bagci} and the measurement results with two receiving antennas, by increasing the number of receivers and using their CSIs simultaneously, the detection performance increases. Methods~2 and especially 3 perform significantly better with a single RX antenna (avg. tamper detection rate of 99.6\% at a 0\% false positive rate for method 3) than method 1 \cite{Bagci} for which we used 2 receiving antennas. The superior performance comes at the cost of increased computational complexity. A disadvantage of method~3 is a delay equal to $N_{\mathrm{On}}$ number of frames. By comparing the detection performance of the DCAE with a different number of layers (DCAE1 and DCAE2) in Fig.~\ref{roc}, we find that adding one layer in encoding and decoding only slightly increases the already excellent performance.
\section{Conclusion}
In this letter, we proposed a DCAE-based approach for detecting a physical tamper attack using CSI in an OFDM-based wireless communication system. The main challenge of this problem is to distinguish between antenna orientation changes and communication environment changes. To achieve a robust attack detector, we use a post-processing of the DCAE. In our experiment, we achieved on average a tamper detection rate of up to 99.6\% at a false positive rate of 0\% in two different environments, which outperforms existing work.
\ifCLASSOPTIONcaptionsoff
  \newpage
\fi
\bibliographystyle{ieeetr_noParentheses}
\bibliography{references_DB}
\end{document}